# A Robust Certificate Management System to Prevent Evil Twin Attacks in IEEE 802.11 Networks

Yousri Daldoul
Faculty of Sciences of Monastir, University of Monastir, Monastir, Tunisia
yousri.daldoul@fsm.rnu.tn

*Abstract-The* evil twin attack is a major security threat to WLANs. An evil twin is a rogue AP installed by a malicious user to impersonate legitimate APs. It intends to attract victims in order to intercept their credentials, to steal their sensitive information, to eavesdrop on their data, etc. In this paper, we study the security mechanisms of wireless networks and we introduce the different authentication methods, including 802.1X authentication. We show that 802.1X has improved security through the use of digital certificates but does not define any practical technique for the user to check the network certificate. Therefore, it remains vulnerable to the evil twin attack. To repair this vulnerability, we introduce Robust Certificate Management System (RCMS) which takes advantage of the digital certificates of 802.1X to protect the users against rogue APs. RCMS defines a new verification code to allow the user device to check the network certificate. This practical verification combined with the reliability of digital certificates provides a perfect protection against rogue APs. RCMS requires a small software update on the user terminal and does not need any modification of IEEE 802.11. It has a significant flexibility since trusting a single AP is enough to trust all the APs of the extended network. This allows the administrators to extend their networks easily without the need to update any database of trusted APs on the user devices.

1. INTRODUCTION

IEEE 802.11 [1] networks are widely used thanks to their high throughput capacity and easy installation. Due to the broadcast nature of these networks, any attacker can eavesdrop on their transmitted data. Therefore, WLANs must provide enough security to protect the user privacy. 802.11i is the principal amendment that intends to improve the security. It defines several protocols and algorithms to provide authentication, integrity and confidentiality services. A WLAN that supports 802.11i is called a Robust Security Network (RSN). Although 802.11i introduces robust mechanisms, an RSN is still vulnerable to several attacks, such as the evil twin attack. The principle of this attack is to install a rogue AP which impersonates a legitimate AP. When a new user wants to join the WLAN, he may confuse the rogue AP with the legitimate one and associate with the rogue AP. This allows the adversary to perform several attacks, such as intercepting the user credentials, stealing sensitive information and eavesdropping on the victim communication.

WLANs are suitable for multiple environments. They can provide public access to open networks in different areas, such as malls, municipalities, libraries and airports. They can also provide private access to authorized users, like students, employees, customers, hotel guests and family members. This is possible thanks to the different supported authentication methods. In fact, 802.11i defines 3 authentication methods: Open System Authentication (OSA), Pre-Shared Key (PSK) and 802.1X. OSA does not require any password and allows any user to join the network. PSK requires the users and the AP to share the same password. 802.1X requires an authentication server (AS) that authenticates the users by means of their credentials (e.g. username and password). Open networks are vulnerable to evil twin attacks since there is no mutual authentication between the user and the AP. PSK allows mutual authentication using the shared password. As long as the password is protected, the connection is secure and the evil twin attack is impossible. PSK is suitable for small WLANs, such as residential networks, where few users are able to share the password securely with each other. It is not convenient for public or large networks since the attacker is able to obtain the password which allows the rogue AP to succeed the mutual authentication with the victims. On the other hand, 802.1X [2] is suitable for large networks since it provides every user with his own credentials. It allows the user to authenticate the network by means of the digital certificate of the AS, while the user credentials allow the AS to authenticate the users. This authentication method is widely used by companies, hotels, shops and universities, such as the largest university network Eduroam [3]. Unfortunately, the evil twin attacks are easy to perform against 802.1X and allow the attacker to steal the user credentials. This is because the victim ignores the AS certificate and can trust any self-signed certificate provided by the rogue AP. Therefore, he may send his credentials in plaintext to the attacker within a TLS tunnel.

Despite the robust security mechanisms of 802.11i, the evil twin attack is easy to perform. As a result, a large number of studies have been carried out to prevent this attack. However, most of them do not provide a trustworthy detection since they may trust rogue APs and alert from legitimate APs. In addition, several approaches are not practical since they have extensive requirements (e.g. additional hardware, extensive use of the bandwidth, multiple network interfaces, costly signed certificates, etc.). Besides, we notice that all the reviewed proposals do not provide enough security and are easy to bypass. Therefore, it is necessary to define a practical and reliable approach to efficiently prevent the evil twin attacks.

We believe that a robust solution for the evil twin problem must rely on digital certificates. This is because the rogue AP cannot impersonate the legitimate AP without the private key. However, it is essential to provide the user with a practical and reliable method to verify the AS certificate. This allows the secure association with trusted WLANs and the efficient detection of rogue APs.

In this paper, we define a Robust Certificate Management System (RCMS) to prevent all evil twin attacks in WLANs. Our proposal is suitable for both small and large networks using 802.1X authentication. It runs entirely on the user device and does not require any protocol modification. It allows the user to strongly authenticate the AS using an additional code of a limited length, called the verification code. Upon the first association to an SSID, the user is requested to introduce his credentials (e.g. certificate or username/password) and the verification code. Once the AS certificate is checked correctly, the root Certification Authority (CA) of the AS certificate is considered as the trusted CA of the current SSID. Therefore, for any subsequent association to a given SSID, any AS certificate is trusted if its root CA is the trusted CA of the SSID. This allows the network administrators to easily extend their networks and to deploy multiple AS with different certificates issued by the same CA. The user must provide the verification code only if the information stored by RCMS on the user device does not allow trusting the AS (e.g. first association to the SSID or modified public key of the root CA). RCMS efficiently prevents evil twin attacks thanks to the reliable verification of the AS. Besides, our proposal is practical since it only requires slight software updates on the user device.

To summarize, we study the evil twin attacks in WLANs and the limitation of existing security mechanisms. Our main contribution is to introduce a new mechanism, called RCMS, in WLANs employing 802.1X authentication. RCMS allows the reliable check of the AS using a new verification code entered by the user. Therefore, it prevents all evil twin attacks and allows the secure association to legitimate APs.

The remainder of this paper is organized as follows. The next Section introduces related work studying the rogue AP detection in WLANs. Then, Section 3 presents the different authentication methods of 802.11i and their limitations against the evil twin attacks. Section 4 presents the threat model. We introduce RCMS in Section 5 and we conclude in Section 6.

## 2. RELATED WORKS

Extensive research is carried out to define secure protection mechanisms against evil twin attacks. The existing approaches can be classified into 4 main families: traffic anomaly, location, fingerprint and cryptography based approaches.

*2.1 Traffic anomaly based approaches*

A large number of approaches are defined for the case of a rogue AP relaying the communication between the victims and the legitimate AP. Since this type of evil twins increases the number of wireless hops and the delays, the authors of [4] choose the inter-packet arrival time as the detection parameter. In [5], the round trip time (RTT) of ICMP packets is used for the Evil Twin detection. We believe that both methods are not precise as the delays may vary significantly in WLANs due to several factors such as the buffering delays, the used data rates, the number of users and the signal strength. Besides, bridges will be considered as rogue APs since they operate as relays. Another characteristic of evil twins acting as relays is frame forwarding on the wireless channel. This characteristic is considered by the proposal of [6] which continuously monitors the medium to capture and compare the transmitted frames. It classifies APs with frame forwarding as evil twins. Legal AP Finder (LAF) [7] is a similar approach which relies on the frame forwarding behavior of the rogue APs. Instead of comparing all data frames, it only examines the TCP 3-way handshake packets. We note that both [6] and [7] have significant drawbacks and limited accuracy. For example, they are not suitable for encrypted networks because the encryption algorithm makes the forwarded frames different from the original frames. In [8], the rogue AP detection relies on the statistics of the data transmitted by the different APs. An evil twin is identified if it transmits the same amount of data than another AP during the same time interval. A similar proposal is presented in [9] and detects the forwarding behavior by monitoring the arrival time of frames having similar lengths. It requires multiple wireless interfaces (minimum of two) to scan the different channels, and necessitates a long scan period to detect any forwarding behavior. PrAP-Hunter [10] is a detection mechanism for network administrators. It operates on a dedicated device with two wireless interfaces. The first interface associates with an AP and transmits data, while the second interface interferes with channels 1 to 11 sequentially. If the first interface notices throughput degradation when the second interface is interfering with specific channels, this indicates that the AP is an evil twin forwarding data. It is clear that this proposal suffers from significant drawbacks; not only does it waste the bandwidth, but also it does not provide a trustworthy detection. This is because the throughput of WLANs is variable due to several factors, such as medium sharing, interference with legitimate devices, collisions and channel fading. Therefore, PrAP-Hunter cannot ensure any effective protection against rogue APs.

Other approaches consider the case of rogue APs having their own Internet connection. In [11], the detection method relies on the principle that the different APs of a single Extended Service Set (ESS) usually use the same gateway. Therefore, it verifies the gateways of the visible APs belonging to the same SSID, and detects the presence of a rogue AP if different gateways are used. Similarly, Rogue AP Finder (RAF) [12] compares the transmission paths to a given server over the different APs of a particular ESS. It reports the presence of a rogue AP if different paths are used. These methods may work if both the rogue AP and the legitimate AP are visible. Otherwise, the attacker cannot be detected. Besides, the two proposals may detect the presence of an evil twin but cannot distinguish between rogue and legitimate APs.

Several other approaches consider both types of evil twins: relays and those having their own gateways. In [13], the authors combine the gateway check with the frame forwarding verification. BiRe is another detection mechanism defined in [14]. It requires two wireless interfaces associated with two different APs. Every interface sends a TCP SYN packet to a particular server which acknowledges the other interface. The absence of an acknowledgement indicates that one of the two APs is an evil twin. We note that BiRe cannot detect the attack if only one AP is available. Besides, it has excessive requirements which make it impractical. The proposal of [15] intends to alert the network administrator of any existing evil twin. It uses a sniffer that captures and analyses the transmitted frames. It considers that the attacker necessarily sends deauthentication frames to disconnect the victims from legal networks and connect them with the rogue AP. Therefore, an attack is detected if excessive Association Response frames are intercepted. Unfortunately, this proposal is not suitable for many types of evil twins. EvilScout [16] is defined for a very specific case of evil twins when the rogue AP operates on the same channel of the legitimate AP and impersonates the MAC address of the legitimate AP. In this case, the detection is based on anomalies related to MAC address conflicts.

*2.2 Location-based approaches*

To prevent the evil twin attack, the authors of [17] suggest the connection of the legitimate AP to a display that confirms the user connection to the right network. This proposal requires an additional device for every AP and a line of sight between the users and the display. This solution is expensive and not suitable for large networks since the simultaneous verification of multiple people using a single screen is not practical.

The principle of crowd sensing is used by CRAD in [18]. The crowd is composed of the mobile users connected to a specific ESS. Every user device should profile the different available APs by recording their signal strengths (i.e. RSSI) over time. Then it shares its measurement reports with the other members of the crowd. The ratio of reports containing a significant variation of the RSSI is used as an indicator of an existing rogue AP. However, an attacker can broadcast forged reports to decrease the ratio of reports with RSSI anomalies. A similar approach based on crowd sensing is proposed in [19]. It uses the CSI (Channel State Information) and AoA (Angle of Arrival) to improve the detection accuracy, and detects the attack if the spatial location of the AP changes. Another proposal based on RSSI is defined for residential networks in [20]. It considers that the signal strength is a stable parameter that can be used to detect evil twins. This assumption is not valid due to the user mobility and cannot provide reliable and precise attack detection. The principle of RSSI monitoring is also used in [21] to detect rogue APs based on their location. However, instead of using the crowd collaboration, the authors suggest to install multiple sensors that record the RSSI evolution. These measurements are transmitted to and processed by a remote server to detect any anomaly. This proposal alerts the network administrator if a rogue AP is detected but does not prevent the attack.

In [22], the authors use the principle of "trust by location" that records all the visible APs upon the first association to an AP. For subsequent connections, the AP is trusted if the variation of the neighbor networks does not exceed a given threshold. Otherwise, it is classified as an evil twin. In [23], the detection system starts by classifying all the visible APs as authorized and records their parameters in a white-list. Then, it checks for any suspicious modification of different parameters to report a rogue AP. For example, if a new AP is detected after the initialization step, it is considered as an evil twin. It is clear that this approach is not reliable as it may classify many legitimate APs as illegal and may trust rogue APs.

*2.3 Fingerprint-based approaches*

ETGuard [24] is an administrator-side mechanism which detects rogue APs based on the recorded fingerprints. It runs on a dedicated server and continuously records the beacon frames. Since the fingerprints are calculated from the beacon frames, any attacker is able to spoof these frames and impersonate legitimate APs. This affects the reliability of ETGuard. Multiple approaches use radiometric signature as a unique identifier of each device. In [25], the observation of the clock skew is used as the AP fingerprint. In [26], the authors use the CSI to extract the physical layer information of the transmitter. They consider that the phase errors depend on the device and can be used to create a unique fingerprint of any AP. Another approach [27] extracts the AP fingerprints from the power amplifier and frame distribution of the received data. The mechanism proposed in [28] detects rogue APs based on multiple parameters, namely the clock skew, the used channel, the received signal strength and the beacon transmission duration. These proposals must be initialized with a fingerprint list of authorized devices. Due to this constraint, any network extension or modification requires the update of the fingerprint list of every user. We note that the attacker can obtain a device identical to the used AP. This allows the rogue AP to produce the same fingerprint and to bypass the detector.

*2.4 Cryptography-based approaches*

VOUCH-AP [29] is among a few proposals that use digital certificates to authenticate the legitimate AP and to prevent the attacks. The authors provide each AP with a certificate issued by a trusted Certification Authority (CA). This certificate includes the network SSID and aims to prevent WLAN impersonation. Unfortunately, the SSID is not a unique identifier for WLANs and can be used by different networks simultaneously. Therefore, an attacker can obtain a signed certificate from a trusted CA for any SSID and perform the evil twin attack. As a result, this proposal is not secure enough and incurs additional costs related to the purchase of a signed certificate for every AP.

In [30], the authors show that the use of WPA2-Enterprise (i.e. 802.1X authentication) remains vulnerable to the evil twin attack which allows the adversary to steal user credentials. This is because the user ignores the AS certificate and cannot check it. Therefore, he may accept any certificate, including that of the attacker. To solve this problem, the authors suggest to display WPA2-Enterprise networks in a list of pairs « **SSID, AS name** ». As the authors recognize, this solution is not secure since the attacker is able to produce a certificate (either self-signed or signed by a trusted CA) containing the same AS name. In [31], the authors show that the 802.1X authentication used in Eduroam networks does not sufficiently secure WLANs since most users do not check the AS certificate. Therefore, they suggest activating the check of the AS name (i.e. displaying the information of the AS certificate in an interface and asking for the user permission before pursuing the authentication). This is not a reliable approach since self-signed certificates are widely used in WLANs, allowing the attacker to use any AS name. In addition, most users are not aware about the AS name and trust the WLAN based on its SSID. A similar study of the Eduroam security [32] shows that the authors were able to access the user credentials of 61% of the tested devices which accepted to associate with a rogue AP. To prevent the evil twin attacks, Eduroam provides a Configuration Assistant Tool (CAT) [33] that configures the user device with the Eduroam network profile. This solution requires the user to download and execute CAT. Since the use of CAT is not mandatory, most users may ignore it. We note that the created profile does not prevent the association with a rogue AP but allows to inform the user that the network details have changed and requests the user authorization to pursue the authentication. Therefore, we believe that CAT is neither practical nor reliable.

### 3. BACKGROUND OF WLAN SECURITY

*3.1 Network Discovery*

In a WLAN, every AP is identified using a unique identifier called BSSID (i.e. the MAC address of the AP). To extend the coverage of a WLAN, the administrator may install multiple APs. The extended network is called Extended Service Set (ESS) and is identified using a string called SSID. To join a WLAN, the user station (STA) follows 3 steps: network scanning (active or passive), authentication and association. During the first step, the STA scans the different channels of the spectrum to find the available networks. Using passive scanning, the STA receives the beacon frames of the visible APs. These frames are broadcasted periodically and contain all the information about the AP, such as SSID, BSSID and the security protocol. They allow the user to select the desired SSID. If multiple APs belonging to the same SSID are visible, the STA selects the AP with the highest signal strength (i.e. RSSI) as it is expected to provide the highest throughput. During the user mobility, the STA may perform a seamless handover from one AP to another within the same ESS. This handover is defined by the IEEE 802.11 standard and does not require the user permission.

*3.2 Authentication and Association*

The second step after network scanning is user authentication. Current networks support 3 authentication methods: Open System Authentication (OSA), Pre-Shared Key (PSK) and 802.1X. The first method does not use any password and does not provide any authentication. It allows any user to join the network if his MAC address is not black-listed. This method does not use encryption and the network frames are transmitted in the clear. A recent enhancement of open authentication is defined in [34] and allows data encryption in open networks. As there is no way to authenticate the users and the WLAN, an evil twin attack is easily performed against open networks and cannot be detected or prevented.

The second authentication method is PSK. It requires the users and the AP to share the same password. During authentication, both the STA and the AP must prove knowledge of the secret. This ensures mutual authentication between the user and the network. Without the password, a rogue AP cannot authenticate to the STA and cannot establish a connection with the victim. Since the transmitted frames are encrypted in a WLAN protected with PSK, the adversary cannot eavesdrop on the data and cannot perform any attack. We note that PSK is practical in small WLANs, such as residential networks, as long as the few users are able to keep the password confidential. PSK is not suitable for public or large networks since the password is accessible to any user. For example, several restaurants and cafes provide their customers with free connections to WLANs protected by PSK. Typically, they provide them with the password within the receipt. This allows any malicious user to obtain the secret, to impersonate the legitimate AP and to perform the evil twin attack.

The third authentication method is 802.1X. It is widely known as WPA2-Enterprise. It requires an authentication server (AS) which performs the mutual authentication with the users by the intermediate of the AP. The AS uses its certificate to authenticate itself to the user. If the user trusts the AS certificate, he uses his credentials (e.g. certificate or username/password) to authenticate to the server. The AS certificate allows the establishment of a secure tunnel between the user and the AS to perform the user authentication. 802.1X authentication is suitable for large networks since every user has his own credentials and can identify legitimate APs thanks to the AS certificate. It prevents evil twin attacks and guarantees data confidentiality in public networks, even if the same username and password are publicly shared, as long as the users are able to verify the AS certificate. It can also be used in small networks since many commercial APs have integrated AS and can use self-signed certificates. Unfortunately, a large number of evil twin attacks are successfully achieved against 802.1X and allow the adversary to steal the user credentials and to eavesdrop on the traffic. This is because most users cannot verify the AS certificate and accept to authenticate with rogue APs providing self-signed certificates. Therefore, they establish a secure tunnel with the attacker who becomes able to perform multiple attacks.

As aforementioned, OSA is a null authentication protocol. It uses a two-frame exchange. The first frame contains the STA identity (i.e. the MAC address) and requests authentication. The second frame returns the authentication result. If the result is "successful," the STA and the AP are considered mutually authenticated. As depicted in Figure 1, the authentication step of PSK is either OSA or Simultaneous Authentication of Equals (SAE). SAE intends to make PSK resistant to offline dictionary attacks. It generally uses the elliptic curve cryptography to derive an intermediate key, called Pairwise Master Key (PMK), from the PSK. When OSA is used with PSK, PMK is identical to the pre-shared key (i.e. PMK=PSK). In the case of 802.1X, the authentication step relies on OSA.

The association step is a two-frame transaction sequence following the authentication. It is initiated by the STA and allows the negotiation of the connection parameters. In the case of open authentication, no more steps are required and the user device is successfully associated to the WLAN. But all the frames of open networks are transmitted in the clear. If 802.1X is used, the 802.1X authentication step starts following the association. It allows the STA and the AS to derive the PMK from the TLS master key which is used to establish the TLS tunnel. Then, the AS sends the PMK to the AP. This allows the STA and the AP to share the same key.

If PSK or 802.1X is used, the AP and the STA start the 4-way handshake to derive the Pairwise Transient Key (PTK) from the PMK. This handshake intends to confirm that a live STA holds the PMK and to derive a fresh PTK. The PTK is used to encrypt the transmitted data frames. Figure 1 illustrates the network access steps using the 3 authentication methods.

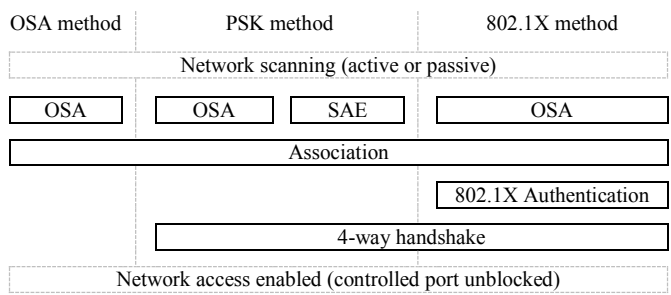

Figure 1. Network access steps in IEEE 802.11 WLANs

*3.3 802.1X authentication*

802.1X authentication supports different credential types, such as digital certificates, usernames and passwords, secure tokens, and mobile network credentials (i.e. GSM and UMTS secrets). A WLAN employing 802.1X typically consists of user devices, one or multiple APs belonging to the same ESS, and one AS. For large networks, such as Eduroam [3], it is possible to use multiple servers. Figure 2 illustrates a simple WLAN using 802.1X. The most used authentication server is the RADIUS server which uses the RADIUS protocol to communicate with the AP. Therefore, the mutual authentication between the user and the AS is performed using two protocols: EAP over LAN (EAPOL) and RADIUS. EAPOL is introduced by 802.1X and relies on EAP [35]. It defines additional frames to support wired and wireless LANs. EAP is an authentication protocol used between the STA and the AS. The EAP messages are transmitted within 802.11 frames over the wireless medium, and within RADIUS packets between the AP and the AS.

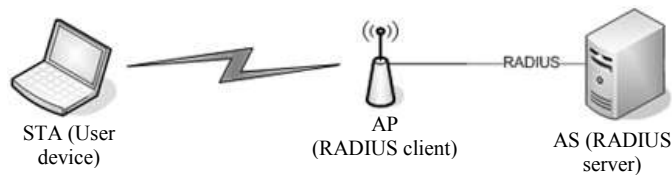

Figure 2. Network architecture using 802.1X authentication

EAP supports multiple authentication methods [36] which can be classified into two principal categories: password-based and TLS-based methods. However, not all of them are compliant with WLANs. In fact, 802.11i requires the use of an EAP method capable of generating the keying material [37]. Therefore, only TLS-based methods are compliant with the RSN requirements. They establish a secure TLS tunnel between the user device and the AS using the server certificate. If the user authenticates using his username and password, a password-based authentication method must be used through the encrypted tunnel and is called inner or tunneled method. This inner method may be EAP or non-EAP method, depending on the used TLS-based EAP method.

The most popular TLS-based EAP methods are:

- EAP-TLS [38]: This method allows the user and the AS to mutually authenticate using certificates. Therefore, both the AS and the user must have certificates. This method is mainly used in large companies where the network administrators take care of configuring the device of every employee individually. We note that EAP-TLS is supported by all devices since it is among the requirements of WPA2.
- EAP-TTLS [39]: This method only requires the server to hold a certificate and is, therefore, more practical than EAP-TLS. It allows the user to authenticate using his username and password through the TLS tunnel. This method supports multiple inner methods. It supports both non-EAP (e.g. PAP, CHAP and MSCHAPv2) and EAP methods (e.g. EAP-MD5 and EAP-MSCHAPv2). Like EAP-TLS, EAP-TTLS is also supported by any device compliant with WPA2.
- PEAP [40]: This method is similar to EAP-TTLS, but only supports EAP methods as inner methods.

On the other hand, the most used inner methods are:

- PAP [41]: This method allows the user authentication using his username and password. These credentials are transmitted in plaintext and are easily accessible if the TLS tunnel is established with the attacker.
- CHAP [42], MS-CHAP [43] and EAP-MD5 [37]: These are one-way authentication methods which allow the server to authenticate the user using challenges.
- MS-CHAPv2 [44] and EAP-MS-CHAPv2 [45]: These methods provide mutual authentication using challenges; both the server and the user must prove their knowledge of the user password.

Figure 3 illustrates an example of 802.1X authentication using EAP-TTLS and EAP-MD5 as the inner method.

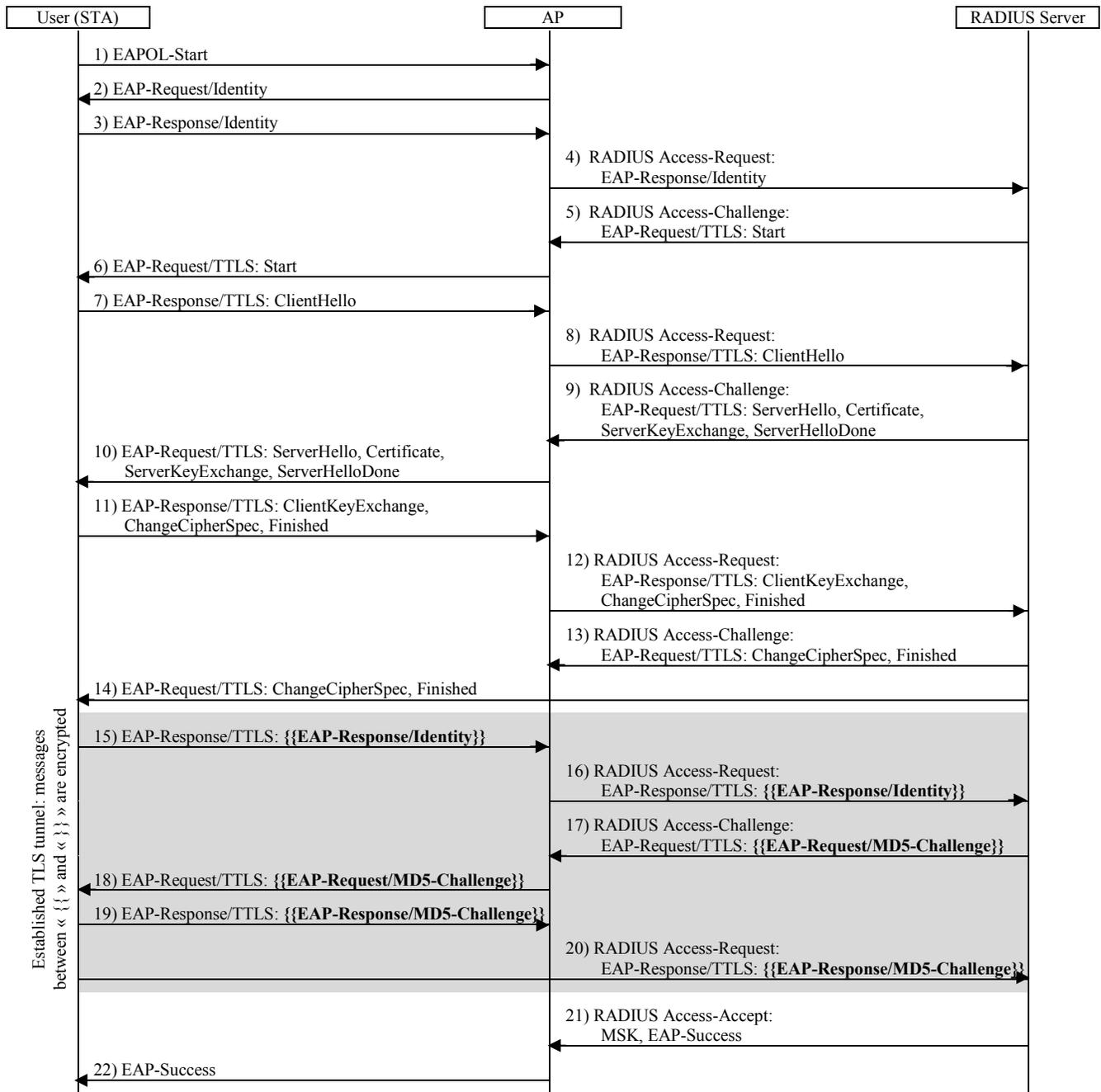

**Figure 3. 802.1X authentication using EAP-TTLS and EAP-MD5**

## 4. THREAT MODEL

### 4.1 Attack objectives

This section introduces the different evil twin attacks against 802.1X authentication according to the adversary expectations. In fact, we distinguish two attack objectives:

- Credential theft
- Data relay (man-in-the-middle)

In the first case, the attacker intends to steal valid credentials in order to access the WLAN as an authorized user. This is a typical attack against several private and paid networks (e.g. university, airport and Internet provider WLANs) where the network access is limited to authorized users only. The damages of this attack depend on the access rights of the victim and vary from simple to very harmful damages, such as bandwidth sharing, paid plan consumption and unauthorized access to personal documents and sensitive information.

The second objective is to relay the victim's data. We note that the attacker can easily provide an Internet connection using different methods, such as mobile, wireless or wired networks. Once the victim is connected to the Internet through the rogue

AP, the adversary can perform various passive and active attacks, such as data eavesdropping and user redirection to phishing websites. Although most sensitive websites use https to encrypt and protect the user data, several websites still use http and can be spied. Since many people use the same username and password to access different accounts, stealing the credentials from unencrypted websites may allow the attacker to gain access to the user accounts of sensitive websites. In addition, phishing websites may succeed to steal sensitive data (e.g. passwords and credit card information) and to convince the victim to download malware. We note that some malware are very harmful and allow the attacker to easily spy and control the victim device.

To steal the user credentials, the attacker does not need to provide an Internet connection or to be in visibility with a legitimate AP. He only needs to install a rogue AP, to capture the required information and to leave. But to relay the user data, the attacker must provide an Internet connection either using his own gateway (mobile network or wired LAN) or using the legitimate WLAN. In the latter case, a legitimate AP must be visible.

*4.2 Credential theft*

As previously mentioned, EAP-TLS allows the user and the AS to mutually authenticate using certificates. This is the most secure EAP method since the user certificate is useless to the adversary without the user's private key which is never transmitted over the network. Therefore, EAP-TLS is perfectly secure against credential theft. For the other TLS-based methods, the credentials are safe if the user associates with the legitimate AP. But if the victim associates with the rogue AP and accepts any certificate, the encrypted tunnel is established with the attacker who can decrypt the tunneled data. In this case, the attacker can obtain the victim credentials using two possible attacks: downgrade and dictionary attacks.

The downgrade attack allows the adversary to negotiate the weakest possible EAP method in order to facilitate the access to the credentials. In fact, EAP has several methods which are not necessarily supported by every STA and AS. In a typical scenario, the AS suggests EAP methods from strongest to weakest till a method is accepted by the STA. This allows the selection of the strongest method supported by both parties. In the case of a downgrade attack, the malicious AS suggests the methods from weakest to strongest in order to use the weakest possible. If EAP-TTLS with PAP is used, the attacker receives the user credentials in plaintext and no more action is required. But if a challenge-response method is used, the attacker performs an offline dictionary attack using the challenge and the received response. This attack succeeds only if the victim password figures within the dictionary of likely passwords used by the attacker. Since the adversary's purpose is to steal the user credentials, he does not need to succeed the authentication step or to provide the victim with an Internet connection. He only needs the credentials in plaintext or the challenge and the corresponding response.

*4.3 Data relay: Man-In-The-Middle (MITM)*

EAP-TLS is not secure against MITM; if the victim trusts the rogue AP and accepts its certificate, the adversary accepts the victim certificate and succeeds the mutual authentication. In this case, the victim data will be relayed through the rogue AP. Similarly, the other TLS-based EAP methods do not prevent the victim from accepting the attacker certificate. Once the certificate is accepted, the success of the mutual authentication depends on the security of the inner method. Hence, if the adversary succeeds to negotiate an inner method that allows one-way authentication (e.g. PAP, CHAP, EAP-MD5 and EAP-MSCHAP), he succeeds the authentication step easily.

If the victim refuses all weak inner methods and only accepts a mutual authentication method (e.g. EAP-MSCHAPv2), the attacker must prove knowledge of the password and reply correctly to the challenge. This makes the authentication more difficult, but possible if a legitimate AP is visible to the attacker. In this case, the attacker impersonates the STA to the AS and establishes a second tunnel with the AS. Then, he negotiates the same authentication method. Upon the reception of the AS challenge, the attacker sends this challenge to the victim. Then, he forwards the victim's response and challenge to the AS. Finally, he forwards the AS response to the victim. This allows the attacker to succeed the mutual authentication with both STA and AS. In the remainder, the victim and the rogue AP derive the same session keys and the victim data are relayed through the evil twin as desired by the attacker.

In many environments (e.g. restaurants, cafes, libraries, etc.), the network is available for customers and is protected using the same credentials for all the users. Therefore, the attacker is able to obtain the shared password and to succeed the mutual authentication of the inner method without the need to interact with the legitimate AP. In this case, the only protection against the evil twin attack is the verification of the AS certificate.

*4.4 Summary*

To summarize, the evil twin attacks against 802.1X are possible when the victim does not verify the AS certificate and accepts any one. If the user is able to check the certificates and associates with authorized APs, no evil twin attack is possible. Table 1 illustrates the security level of the most used EAP methods and the possible attacks to attain the adversary objectives. For a given inner method, we consider that this method is selected following the downgrade attack and is the weakest possible method that the attacker can negotiate.

**Table 1. Security level of most used EAP methods**

| Main EAP method | Inner method | Adversary objective | |
|---|---|---|---|
| | | Credential theft | Data relay: MITM |
| EAP-TLS | null | Impossible | Easy |
| EAP-TTLS | PAP | Easy | Easy |
| | CHAP, EAP-MD5, MSCHAP, EAP-MSCHAP | Possible using an offline dictionary attack | Easy |
| | MSCHAPv2, EAP-MSCHAPv2 | | Possible: requires legitimate AP visibility |
| PEAP | EAP-MD5, EAP-MSCHAP | | Easy |
| | EAP-MSCHAPv2 | | Possible: requires legitimate AP visibility |

## 5. ROBUST CERTIFICATE MANAGEMENT SYSTEM

In this section, we introduce our approach to protect WLANs against any type of evil twin attack. Our solution is called Robust Certificate Management System (RCMS) and is defined for 802.1X authentication. It allows the user device to easily and precisely check the AS certificate. Therefore, RCMS only accepts legitimate certificates and associates with authorized networks. It rejects any authentication with rogue APs thanks to a new code called "**verification code**". This code allows the verification of the server certificate and the authentication with legitimate networks. **RCMS is suitable for all types of credentials**, but we mainly focus on the case of username/password pairs which are widely used with 802.1X. Upon the first association to an SSID, the user must provide 3 values instead of 2: username, password and the verification code. If the code is valid, the network is trusted and is added to the list of trusted networks. For subsequent associations to trusted APs, the code is not requested unless the public key of the root certificate is modified.

To successfully authenticate the servers, RCMS maintains the certificates of the root CA instead of the AS certificates. When the STA receives a new AS certificate, it checks the root certificate and accepts the AS certificate if the root CA is trusted. This allows large networks to use multiple servers with different certificates having the same root certificate. In the case of a small network having a single AS and a self-signed certificate, the root certificate is the AS certificate. Therefore, **our design is suitable for both small and large networks**. We note that the root CA does not need to be public as this incurs additional fees without any improved security. However, it is possible and more practical (i.e. free and more secure) to use a private CA, i.e. a self-signed certificate that is used directly to sign the AS certificates or to sign intermediate CA certificates.

In addition, RCMS associates a single root certificate to an SSID. Therefore, a trusted CA is only trusted for the corresponding SSID. Since certificates may be renewed or updated, RCMS is able to update the stored root certificate seamlessly as long as the public key has not changed. But if the public key of the trusted CA is modified, the user must provide a new code to check the AS certificate. We note that updating the certificate does not require the modification of its public key unless the private key is compromised. Moreover, the root certificate is generally valid for many years (typically 3 to 20 years). Hence, **the verification code is rarely requested after the very first association to a WLAN**.

*5.1 Verification code calculation*

The verification code is used to check the AS certificate. It can be calculated differently:

1) **The code is derived either from the AS certificate or from the root certificate**: These are two possible options. In both cases, RCMS saves the root certificate as the trusted CA upon the first successful verification. If the code is derived from the AS certificate, the following constraint must be satisfied: the user must authenticate the first time with the AS for which the code is generated. To get rid of this constraint, we can derive the code from the root certificate. This second option is more flexible and allows the first authentication to occur with any AS of the ESS.

2) **The code is calculated either based on the entire certificate or based on the public key**: If the code is calculated based on the entire certificate, a particular authentication failure may occur in the following case: the code is calculated and then the certificate is updated before the first connection of the user. In this case, the authentication fails and the user must obtain a new code. We note that this is a particular case since a certificate is not modified frequently. However, we can avoid this particular case if we calculate the code based on the public key. In fact, the public key does not change during updates and renewals, unless the corresponding private key is compromised.

3) **The code is calculated using either a hash or a keyed-hash function**: It is possible to use a hash function to calculate the code. In this case, the code is common to all users and is easily accessible by an adversary. Therefore, the code must be long enough to be resistant to a brute force attack. The second option is to calculate the code using a cryptographic hash function and the user password. Therefore, the code is not a common value and varies among users. This allows short codes to be more resistant to brute force attacks; since the adversary ignores the code and the password, he cannot generate asymmetric keys where the keyed-hash of the public key is identical to the verification code.

Although multiple options are possible to calculate the code, we choose the most flexible and secure one. Therefore, **we calculate to code using the keyed-hash of the public key of the root CA**. We suggest the use of HMAC-SHA256 which has an output of 32 bytes. Since this is a very long value for the user, we suggest using the first 6 bytes (i.e. 48 bits) as our code. We believe that this length is long enough to authenticate the root CA and is convenient for the user. In addition, we convert the binary value into base64 and we obtain a string of 8 alphanumeric characters, as follows:

```
Code = base64(first48(HMAC-SHA256(password, CA_PubKey)))
```

(1)

We designed RCMS to accept any AS certificate issued by the trusted root CA. Therefore, the network administrators have two options when the private key of any AS is compromised. The first option is to provide an online revocation list which allows the users to check for revoked certificates. This is practical for large networks with many permanent users, such as Eduroam. We note that the management of the certificate revocation list is beyond the scope of this paper. The second option is to update the public key of the root CA. This forces the users to contact the administrator and to request new codes.

*5.2 List of SSID/CA*

RCMS maintains a list of SSIDs and the corresponding trusted root CA (i.e. list of SSID/CA). This list only contains SSIDs employing 802.1X authentication. In this list, the SSID must be unique but not the root CA. This means that an SSID must have a unique root CA, but a root CA may be associated to multiple SSIDs. This allows the network administrators to use the same root CA with different SSIDs having similar meanings, such as "*University of Monastir 1*" and "*University of Monastir 2*". The list is updated in the following cases:

1) First authentication with an SSID: a new entry is added to the list upon the successful certificate check using the verification code.
2) The root certificate has changed, including the public key: if the user has the right verification code, the existing entry is updated. Otherwise, this is a possible evil twin attack and the authentication must be canceled. If this AP is legitimate, the user must contact the administrator to obtain the new verification code.
3) The root certificate is modified but the public key has not changed: the root certificate is seamlessly updated in the list of SSID/CA and no verification code is requested.

Table 2 depicts an example of the list of SSID/CA. It includes the columns SSID, the public key of the trusted CA, the root certificate fingerprint (allows the detection of any update in the certificate) and the root certificate path (the storage path of the certificate on the user device). It is possible to include addition columns to this list in order to provide more details, such as the date of first association, the update history, etc. Similar to the operating mode of current user devices, it is necessary to store the user credentials in order to provide seamless authentication to trusted networks. These credentials can be saved either in this list or in a separate encrypted list.

**Table 2. List of SSIDs and the corresponding trusted CA (SSID/CA)**

| SSID | Public Key of Root CA | Root certificate fingerprint | Root certificate path |
|---|---|---|---|
| Univ_Monastir | a5f716e894c6… | 487ed1fb85c3… | rootca/univ_m.pem |
| Hotel_SBM | 3749f2ae752b… | 3cb5d0e5edbd… | rootca/hsbm.pem |
| ⋮ | ⋮ | ⋮ | ⋮ |

*5.3 AS certificate check*

The successful check of the verification code means that the root CA is trustworthy. Therefore, the user can accept any certificate issued by this CA. To perform the code verification, the STA must receive the entire certificate chain. This chain includes the AS certificate, the root certificate and any intermediate certificate. It is received during the establishment of the TLS tunnel within the "Certificate" field of the TLS message, as depicted in message 9 of Figure 3. Upon the reception of the certificate chain, RCMS checks the validity of this chain by inspecting the different issuers. Then, RCMS checks if the public key of the root CA exists in the list of SSID/CA and corresponds to the current SSID. If yes, the code verification is not required since the root CA is already trusted. Otherwise, an intermediate code is calculated based on the user password and the public key of the root CA, according to Equation 1. If this intermediate code is identical to the verification code, the root CA is trusted and the list of SSID/CA is updated. Otherwise, the certificate verification fails and the 802.1X authentication is canceled. To summarize, **the AS certificate is accepted only if the certificate chain is valid and the root CA is trusted**.

*5.4 Operating mode*

On the network side, the administrator must configure the network to use 802.1X authentication. For small networks composed of one AP, it is possible to use the internal RADIUS server as the AS. This allows the wireless router to operate as both AP and AS without the need for an additional machine. For an ESS composed of several APs, a single AS is generally enough. If only one AS is used, a self-signed certificate is sufficient. For large networks requiring multiple authentication servers, the administrator should create a private CA with a self-signed certificate to be used as the root CA of the different AS certificates. For very large networks, it is possible to create additional intermediate CAs. Furthermore, the administrator must provide every AS with the entire certificate chain. In the case of Freeradius [46], the certificate chain is typically located in **/etc/freeradius/certs**.

When a new user arrives, he must contact the network administrator to obtain the connection credentials and the verification code. The credentials can be generated automatically using software and printed as part of a document (e.g. student subscription document, hotel bill, cafe receipt, etc.) or transmitted by phone to the user. It is also possible to allow the user to choose his username/password and to enter them manually into a software interface. In both cases (automatic or manual generation), the software must use the root certificate as input and must generate the verification code from the user password and the public key of the root CA, as explained in Equation 1.

The operating mode of RCMS is illustrated in Figure 4. At the beginning, the STA starts the authentication (OSA or SAE) and the association steps with a given SSID, say SSID1. Then RCMS checks if this SSID uses 802.1X authentication or not. If not, there are two possible results: 1) SSID1 does not exist in the list of SSID/CA: in this case, 802.1X authentication is not required (as shown in Figure 1) and the user authentication is successful. 2) SSID1 exists in the list of SSID/CA: in this case, the authentication is rejected. In fact, this scenario corresponds to a WLAN that had used 802.1X authentication and now uses another security policy. Even if this scenario is legitimate, it is not typical. However, it may hide an evil twin attack where the attacker impersonates the SSID and offers an open access to his WLAN. Once the victim is connected, he is redirected to a fake captive portal [47] requesting the user credentials of the legitimate SSID1. For security purpose, RCMS rejects the authentication with an SSID that replaces the 802.1X authentication with another authentication method. Two additional output results of RCMS are depicted in bold in parallelograms 3 and 4. The third output is "Authentication Canceled" and corresponds to the user canceling the authentication. The last output is "Authentication Success".

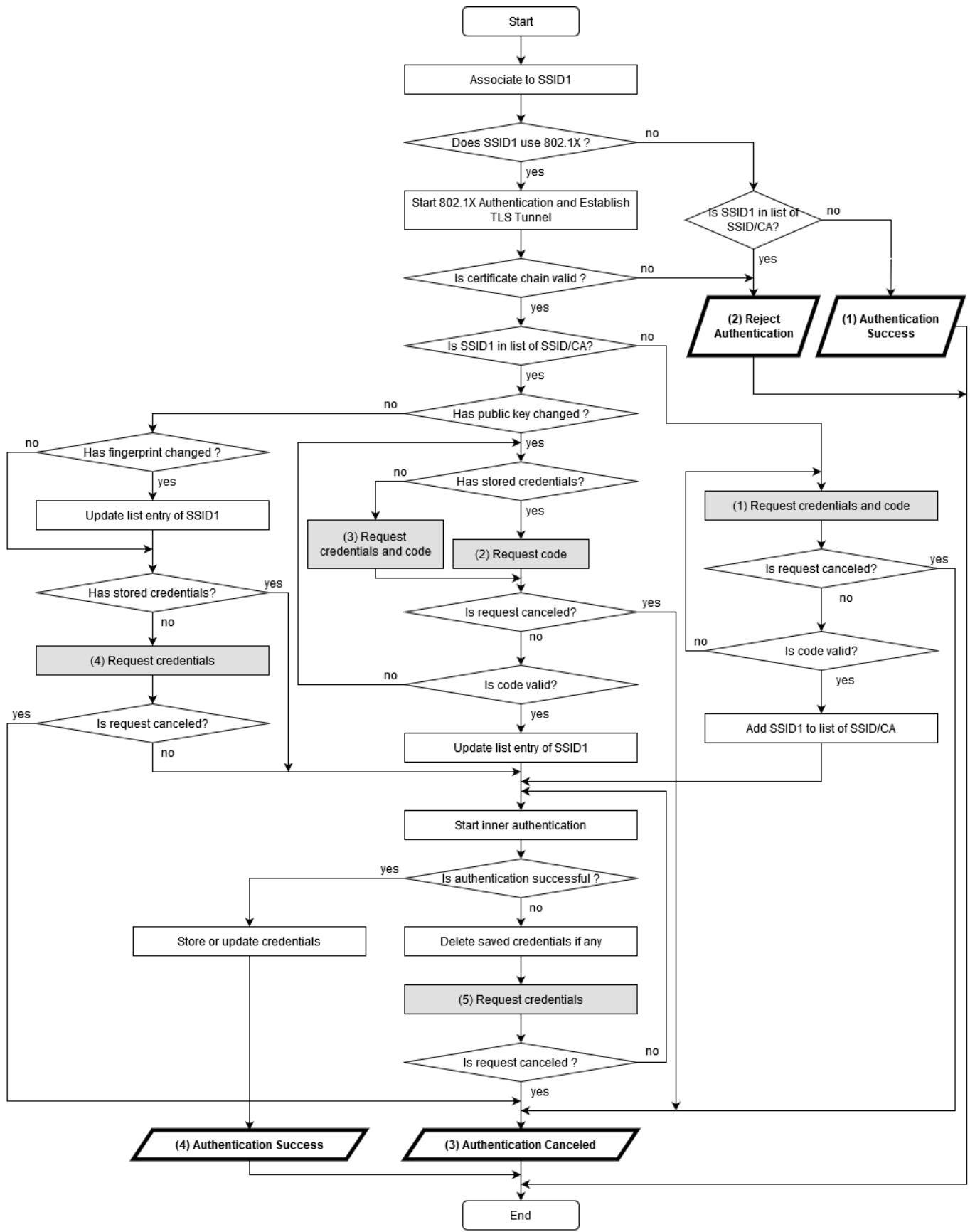

**Figure 4. User Authentication using RCMS**

RCMS requires the user input in 5 situations illustrated in 5 grey rectangles numbered from 1 to 5. The first case occurs when the user associates with the WLAN for the first time. Therefore, he must provide his credentials and the verification code. The second case occurs when RCMS notices that the public key of the root CA has changed compared to the stored value. Since RCMS already has the user credentials, it only requests the new code. The third and fourth scenarios occur when the root certificate was validated during a previous authentication but RCMS has no stored credentials. They may occur when the user provides a wrong username but correct password and verification code during the first authentication. This allows RCMS to add the SSID and the root CA to the list of SSID/CA but does not allow the successful authentication and the storage of the credentials. In the third case, RCMS requests the user credentials and the verification code since the public key of the root CA has changed. However, the verification code is not requested in case 4. The fifth case occurs when the credentials are incorrect, have been renewed or have expired.

## 6. Conclusion

In this paper we studied the evil twin attacks and we showed that the adversary is able to impersonate legitimate networks easily. We explained that the only reliable way to detect and avoid rogue APs is to use digital certificates. Since 802.1X authentication does not define any practical technique to verify the AS certificates, we defined RCMS to identify legitimate networks and to prevent evil twin attacks. RCMS introduces a new verification code which allows the user device to check the AS certificates. Therefore, our proposal allows an easy and practical verification of the network identity. In addition, RCMS runs entirely on the user device and is perfectly compliant with IEEE 802.11 standard. Thus, it only requires software updates to protect the user privacy.

## Funding declaration

This work did not receive any funding from any organization.

## Conflict of interest

The author declares that he has no known competing financial interests or personal relationships that could have appeared to influence the work reported in this paper.

## Author Contribution

Yousri Daldoul wrote and reviewed the manuscript.